\documentclass[12pt]{article}
\newcommand{\be}{\begin{equation}}
\newcommand{\ee}{\end{equation}}
\newcommand{\bea}{\begin{eqnarray}}
\newcommand{\eea}{\end{eqnarray}}
\newcommand{\nn}{\nonumber}

\newcommand{\bv}{{\mathbf v}}
\newcommand{\bB}{{\mathbf B}}
\newcommand{\bem}{{\mathbf e}}
\newcommand{\nl}{\nonumber \\}
\usepackage{color}
 \usepackage{graphics}
\usepackage{subfig}
 \usepackage{epsfig}
  \usepackage{enumerate}
  \usepackage{mathrsfs}
\usepackage{amsfonts}
\usepackage{amssymb}
\usepackage{subfig}
\usepackage{makeidx}
\usepackage{float}
\usepackage{soul}
\usepackage{fancyhdr}
\usepackage{slashed}
\usepackage[toc,page]{appendix}
\usepackage{afterpage}
\usepackage{epstopdf}
\usepackage{array}
\usepackage{booktabs}
\usepackage{multirow}

\begin{document}

\begin{center}
{\Large {\bf Axisymmetric equilibria  with central current  reversal caused  by non-parallel plasma flow
}} \vspace{0.5cm}

A. Kuiroukidis$^{1}$ and G. N. Throumoulopoulos$^{2}$

$^{1}$ Technological Education Institute of Central Macedonia, \\ GR
621 24 Serres, Greece
\\
$^{2}$ University of Ioannina, Department of  Physics, \\Section of
Astrogeophysics, GR 451 10 Ioannina, Greece

Emails: kouirouki@astro.auth.gr,$\; \; $gthroum@uoi.gr

\end{center}

\begin{abstract}

We obtain analytic solutions of a generalised Grad-Shafranov
equation describing steady states   with incompressible plasma
flow of arbitrary direction,   toroidal current reversal
and either nested or non-nested magnetic surfaces. It turns out that the component of the flow velocity non parallel to the magnetic field
  can  result  in normal
equilibria with central current-reversal, i.e. equilibria with nested magnetic surfaces and
monotonically varying pressure profiles.

\end{abstract}

\section{Introduction}
Advanced confinement in  tokamaks is related to Internal Transport
Barriers (ITBs) of energy and particle transport (e.g. see the
review paper \cite{wo}). Experimental evidence indicates that
reversed magnetic shear and sheared flow play a role in the
establishment of ITBs; specifically according to experimental
results of JET \cite{devr} and DIII-D \cite{sh}, on the one hand
the reversed magnetic helps in triggering the ITBs development
while on the other hand the sheared rotation has an impact on the
subsequent growth and allows the formation of strong ITBs. In
certain tokamak equilibria with  ITBs known as current holes, the
central current is nearly zero (e.g. see the review paper
\cite{fu}). According to current hole experiments in JET
\cite{hawk} and JT-60U \cite{fuji} the core current density is
clamped at zero, indicating the existence of a physical mechanism
which prevents it from becoming negative. These observations
motivated a number of theoretical studies on the existence of
equilibria with current reversal \cite{huhe}-\cite{bra}. The
conclusion
is that central current reversal in normal equilibria, i.e. global
equilibria with nested magnetic surfaces and monotonically varying
pressure, can not exist \cite{huhe}-{\cite{brja}.  Potential
physical mechanisms for that prevention are the influence of a
resistive kink magnetohydrodynamic instability \cite{huhe} and
reconnection \cite{brja}. However,
 current-reversal equilibria with non nested
 magnetic surfaces, i.e. having magnetic lobes,
 can exist \cite{robi2012}-\cite{bash} in
 consistence with experimental evidence in tokamaks \cite{mihi}-\cite{lilu}.

The present study was partly motivated by the paper of Ref.
\cite{po},  according to which toroidal rotation opens up the
possibility of normal equilibria with current reversal in the vicinity of the magnetic axis. The
purpose of our study is twofold: first to construct axisymmetric
equilibria with current reversal and incompressible flow with
either nested or non-nested magnetic surfaces,  and second to examine
the existence of normal stationary equilibria with current reversal near the magnetic axis.
The construction is based on a generalized Grad-Shafranov (GGS)
equation with incompressible flow (Eq. (\ref{GGS}) below)
involving five free surface-quantity terms. Specifically a
linearised form of that equation is  solved analytically.

The GGS equation is briefly reviewed in section 2. Solutions with
 current density reversal are derived for a toroidal plasma
of arbitrary aspect ratio
 and rectangular poloidal  cross section with ITER-like
 shaping  in section 3. Part of the construction regarding
 two independent solutions of the respective homogeneous
  partial differential equation is given in Appendix.
Section 4 summarizes the conclusions.
\section{Generalized Grad-Shafranov equation}
We consider the generalized Grad-Shafranov (GGS) equation with
incompressible flow \cite{tass}-\cite{thtapo}: \bea (1-M_p^2)
\Delta^\star \psi -
         \frac{1}{2}(M_p^2)^\prime |\nabla \psi|^2
                     + \frac{1}{2}\left(\frac{X^2}{1-M_p^2}\right)^\prime & & \nl
+\mu_0 R^2 P_s^\prime + \mu_0 \frac{R^4}{2}\left[
\frac{\varrho(\Phi^\prime)^2}{1-M_p^2}\right]^\prime
    = 0 &&
                            \label{GGS}
 \eea
Here,
   the
  poloidal magnetic flux function $\psi(R,z)$  labels the magnetic surfaces,
  where  ($R,\phi, z$) are cylindrical coordinates with $z$ corresponding to the axis of symmetry;
 $M_p(\psi)$ is
 the   Mach function of the poloidal fluid velocity with respect to the
poloidal  Alfv\'en velocity;
 $X(\psi)$ relates to the toroidal magnetic
 field, $B_\phi=I/R$,  through $I=X/(1-M^{2})$; $\Phi(\psi)$ is the electrostatic potential;
 for vanishing flow the surface function $P_s(\psi)$
  coincides with the pressure; $B$ is the magnetic field modulus
  which can be expressed in terms of surface functions and $R$;  $\Delta^\star=R^2\nabla\cdot(\nabla/R^2)$;
  and the prime denotes derivatives  with respect to $\psi$.
  Because of incompressibility the density $\varrho(\psi)$ is also a surface quantity and the
   Bernoulli equation for the pressure decouples from (\ref{GGS}):
\be
 P=P_s(\psi) -  \varrho\left[\frac{v^2}{2}-\frac{R^2(\Phi^\prime)^2}{1-M_p^2} \right]
                          \label{pres}
 \ee
where $v$ is the velocity modulus. Also, $\bv$ can be decomposed in a component parallel and  another non-parallel to $\bB$ as
\be
\bv =\frac{M_p}{\sqrt{\varrho}} \bB -R\Phi^\prime\bem_\phi
                             \label{decomp}
\ee
The  quantities $M_p(\psi)$,
$X(\psi)$, $P_s(\psi)$, $\varrho(\psi)$ and $\Phi(\psi)$ are free
functions. Derivation of  (\ref{GGS}) and (\ref{pres}) is
 provided  in  \cite{tass}-\cite{thtapo}.

Eq. (\ref{GGS}) can be  simplified  by  the  transformation
\begin{equation}
u(\psi) = \int_{0}^{\psi}\left\lbrack 1 -
M_p^{2}(f)\right\rbrack^{1/2} df
                                            \label{trans}
\end{equation}
under which  (\ref{GGS})  becomes \bea \label{gs1} \Delta^{*}u
+\frac{1}{2} \frac{dI^{2}}{d\psi}+\mu_{0}R^{2}\frac{dP_{s}}{du}
+\frac{\mu_{0}}{2}R^{4}\frac{dG}{du}=0
                                  \label{ggs1}
\eea where $G(u):=\varrho(\psi)(d\Phi(u)/du)^2$.
 Note  that no quadratic term as $|{\bf\nabla}u|^{2}$ appears
any more in (\ref{gs1}). Once a solution of (\ref{gs1}) is
obtained, the equilibrium can be completely  constructed with
calculations in the $u$-space.

Instead of (\ref{ggs1}) we will employ the GGS in the following
normalized form
\bea \label{gsh1}
[\partial_{\rho\rho}-(1/\rho)\partial_{\rho}+\partial_{\zeta\zeta}]\tilde{u}
&+&\frac{1}{2}\frac{d}{d\tilde{u}}
\left[\frac{\tilde{X}^{2}}{1-M_{p}^{2}}\right]+\rho^{2}
\frac{d\tilde{P}_{s}}{d\tilde{u}}+\nn \\
&+&\frac{1}{2}\rho^{4}\frac{d\tilde{G}}{d\tilde{u}}=0 \eea Here,
$\rho:=R/R_{0}$, $\zeta:=z/R_{0}$, $\tilde{u}:=u/u_{0}$ (where
$R_0$ is the major radius of the toroidal configuration),
$\tilde{P}_{s}:=P_{s}/P_{0}$, $\tilde{X}:=X/X_{0}$,
$\tilde{G}:=G/G_{0}$, with ITER-relevant data $P_{0}=1.6\times
10^{5}Pa\simeq 1.58 Atm$, $R_{0}=6.2m$, $a=1.1m$ (where $a$ is the
major radius), $u_{0}^{2}:=P_{0}\mu_{0}R_{0}^{4}$, $u_{0}\simeq
17.23 Wb$, $X_{0}:=u_{0}/R_{0}=2.78 Wb/m$,
$I_{0}:=u_{0}/\mu_{0}R_{0}\simeq 2.21 MA$, $G_{0}=4.16\times
10^{3}Kg/m^{3}sec^{2}$.
Using the above parametric values the  pressure and toroidal
current density are given
by \bea \label{piesn} & &P(kPa)=160\tilde{P}_{s}(\tilde{u})-80
\left[\frac{\tilde{X}^{2}}{1-M_{p}^{2}}\right]\frac{M_{p}^{2}}{\rho^{2}(1-M_{p}^{2})}
-\nn \\&-&80\frac{\rho^{2}}{(1-M_{p}^{2})}\tilde{G}(\tilde{u})+\nn \\
&+&160\frac{1}{(1-M_{p}^{2})^{1/2}}
\left[\frac{\tilde{X}^{2}}{1-M_{p}^{2}}\right]^{1/2}
\left[\frac{M_{p}^{2}}{1-M_{p}^{2}}\right]^{1/2}[\tilde{G}(\tilde{u})]^{1/2}-\nn
\\&-&80\frac{M_{p}^{2}}{1-M_{p}^{2}}[\tilde{u}_{\rho}^{2}+\tilde{u}_{\zeta}^{2}]
+160\rho^{2}\tilde{G}(\tilde{u})
\eea \bea \label{reuma}
 & &J_{tor}(MA/m^{2})=0.057\frac{1}{\sqrt{1-M_{p}^{2}}}\frac{\tilde{\Delta}^{*}\tilde{u}}{\rho}+\nn
\\&+&0.0285\frac{(M_{p}^{2})^{'}}{(1-M_{p}^{2})^{3/2}}
\frac{\tilde{u}_{\rho}^{2}+\tilde{u}_{\zeta}^{2}}{\rho} \eea We
hereafter drop tildes for simplicity and choose the following
forms
for the poloidal current density,  pressure and non-parallel-flow function $G$: 
\bea
&&\frac{1}{2}\frac{d}{du}
\left[\frac{X^{2}}{1-M_{p}^{2}}\right]=\alpha_{1}+\alpha^{2}u
\label{pc1}
\\
&&\frac{dP_{s}}{du}=-\beta_{1}+\alpha^{4}u \label{pc2} \\
&&\frac{dG}{du}=-\gamma_{1} \label{pc3} \eea Eqs.
(\ref{pc1})-(\ref{pc3}) are integrated with zero integration
constants for the last two equations and with integration constant
$\alpha_{0}$ for the first one. Also, following \cite{wayu} we
introduce the new dependent variable  $A=u/\rho$ and adopt the
boundary condition $ u|_{b}=A|_{b}=0 $. Additionally,  in connection with potential flow-caused central reversal of the toroidal current density  the poloidal
Mach function  is assigned by
\bea
\label{mach1}
M_{p}^{2}=M_{0}^{2} u ^n
\eea
where $M_0^2$ and $n$ relate to the maximum value of the poloidal Mach function on the magnetic axis and  the shape of $M_p^2$, respectively. This function becomes maximum  on axis and  vanishes on the boundary; therefore it is
peaked on-axis in association with  a respective  on-axis
momentum  source. Alternatively, we have employed the  off-axis peaked poloidal Mach function
 \bea \label{mach}
M_{p}^{2}=M_{0}^{2}u^{m}(u_{a}-u)^{n} \eea where
 $u_{a}$ is the poloidal
 flux function on the magnetic axis and $M_0^2$ relates to the maximum of $M_p^2$; the parameters $m,n$ determine the shape of $M_p^2$ and the position of its maximum,  i.e. this maximum corresponds to  $m u_{a}/(m+n)$ with a measure of the profile width being $u_{a}/(m+n)$.

 Furthermore, following \cite{wayu}, \cite{ma},  we consider a boundary
of rectangular cross section so that $\rho\in
[1-\epsilon,1+\epsilon]$ and $\zeta\in
[-\kappa\epsilon,\kappa\epsilon]$ where $\epsilon:=a/R_{0}\simeq
0.177$ is the inverse aspect ratio and $\kappa\simeq 1.86$ is the
elongation. Thus the boundary value problem for the GGS equation
assumes the form \bea
[\frac{1}{\rho}\partial_{\rho}(\rho\partial_{\rho})+\partial_{\zeta\zeta}]A
+\left(\alpha^{2}-\frac{1}{\rho^{2}}\right)A+\alpha^{4}\rho^{2}A=\nn \\
=-\frac{\alpha_{1}}{\rho}+\beta_{1}\rho+\frac{1}{4}\gamma_{1}\rho^{3}
\label{gsbp}
\\
A|_{b}=0 \label{bc} \eea
\section{Solutions and equilibrium properties}
We consider first the homogeneous boundary value problem and
boundary conditions corresponding to Eqs. (\ref{gsbp}) and
(\ref{bc}): \bea \label{gsbp1}
[\frac{1}{\rho}\partial_{\rho}(\rho\partial_{\rho})+\partial_{\zeta\zeta}]A
+\left(\lambda^{2}-\frac{1}{\rho^{2}}\right)A+\lambda^{4}\rho^{2}A=0\nn \\
A(1-\epsilon,\zeta)=A(1+\epsilon,\zeta)=0\nn \\
A(\rho,-\kappa\epsilon)=A(\rho,\kappa\epsilon)=0 \eea and set
$A(\rho,\zeta)=U(\rho)Z(\zeta)$. We thus have \bea \label{zeq}
\frac{d^{2}Z}{d\zeta^{2}}+\nu^{2}Z=0\nn \\
Z(-\kappa\epsilon)=Z(\kappa\epsilon)=0 \eea with solution \bea
\label{zeqsol} Z(\zeta)=cos(\nu_{l}\zeta),\; \;
\nu_{l}=(l+1/2)\pi/\kappa\epsilon,
\; \; (l=0,1,2,...)\nn \\
\eea
Furthermore setting
 $\lambda_{nl}^{2}:=\nu_{l}^{2}+\mu_{n}^{2}$,
$(n=1,2,...)$ we obtain for $U(\rho)$
\bea
\rho^{2}\frac{d^{2}U}{d\rho^{2}}+\rho\frac{dU}{d\rho}+
(\mu_{n}^{2}\rho^{2}+\lambda_{nl}^{4}\rho^{4}-1)U=0 & &  \label{bessl} \\
U(1-\epsilon)=U(1+\epsilon)=0   & &       \label{bc1}
\eea
The differential equation   (\ref{bessl}) is solved in the Appendix
and its two independent solutions are hereafter called
$J_{1}(\rho;\mu_{n},\nu_{l})$ and $Y_{1}(\rho;\mu_{n},\nu_{l})$. The eigenfunctions
of Eq.  (\ref{bessl}) are therefore of the form
 \bea \label{eig1}
U_{n,l}(\rho)=c_{n,l}J_{1}(\rho;\mu_{n},\nu_{l})+Y_{1}(\rho;\mu_{n},\nu_{l})
\eea
and implementing the boundary conditions (\ref{bc1})  results in the
eigenvalue equations \bea \label{eig2}
c_{n,l}=-\frac{Y_{1}((1-\epsilon);\mu_{n},\nu_{l})}{J_{1}((1-\epsilon);\mu_{n},\nu_{l})}\nn
\\
-J_{1}((1+\epsilon);\mu_{n},\nu_{l})\frac{Y_{1}((1-\epsilon);\mu_{n},\nu_{l})}{J_{1}((1-\epsilon);\mu_{n},\nu_{l})}
+Y_{1}((1+\epsilon);\mu_{n},\nu_{l})=0
\eea
The last of Eqs. (\ref{eig2})
is solved numerically for $\mu_{1},...,\mu_{10}$ since for
computational purposes we have retained terms of up to $n=10$.
The dependence of the parameters $\mu_{n}$ as a function of $n=1,...,10$ (approximated to vary continuously) for
 $l=0,1,2$ is shown in Fig. \ref{fign1}.
\begin{figure}[ht!]
\centerline{\mbox {\epsfxsize=10.cm \epsfysize=8.cm
\epsfbox{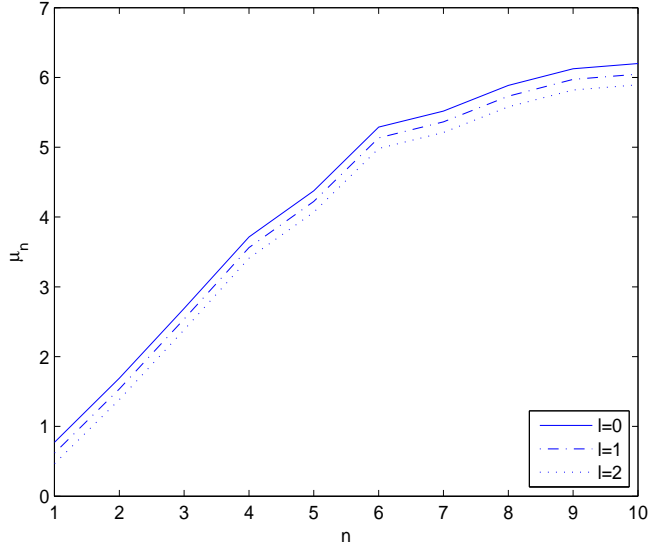}}} \caption[]{The numerically
determined parameters $\mu_{n}$, from Eq. (\ref{eig2}).}
\label{fign1}
\end{figure}
%

To construct a particular solution of the inhomogeneous Eq. (\ref{gsbp}) we
employ the method of expansion
with respect to eigenfunctions \cite{wa,wayu}; accordingly   $A$ is of the form
\begin{equation}
\label{expans}
A(\rho,\zeta)=\sum_{n=1}^{N_{n}}\sum_{l=0}^{N_{l}}A_{nl}U_{n,l}(\rho)cos(\nu_{l}\zeta)
\end{equation}
with the coefficients $A_{nl}$ to be determined. Here the sums can have an arbitrary number of terms
($N_{n}=\infty $, and $N_{l}=\infty $), but for computational purposes we have
retained terms of up to $N_{n}=10$, $N_{l}=4$.
Note that the
functions $U_{n,l}(\rho)$ are {\it not} orthogonal in the interval
$\rho\in[1-\epsilon,1+\epsilon]$.
Also
(\ref{expans}) satisfies the boundary condition automatically.
Then Eq. (\ref{gsbp}) yields
\bea
-\frac{\alpha_{1}}{\rho}+\beta_{1}\rho+\frac{1}{4}\gamma_{1}\rho^{3}=
\sum_{n=1}^{10}\sum_{l=0}^{4}A_{nl}
\left[(\alpha^{2}-\lambda_{n,l}^{2})+(\alpha^{4}-\lambda_{n,l}^{4})\rho^{2}\right]
U_{n,l}(\rho)cos(\nu_{l}\zeta) \nn \\
\label{ggs2}
\eea
Thus, due to the non-orthogonality of the functions $U_{n,l}(\rho)$,
 we multiply Eq. (\ref{ggs2}) by $U_{n^{'},l^{'}}(\rho)cos(\nu_{l^{'}}\zeta)$
and integrate in the relevant intervals. We obtain a linear$-(50\times 50)$
system for the expansion coefficients $A_{n,l}$ which is solved numerically.


Assigning proper values to the free parameters involved in ansatz
(\ref{pc1})-(\ref{pc3}), as given in Table 1, we have constructed
 equilibria with reversed current density. Three examples of them are  shown in Figs.
\ref{fig1a}-\ref{fig3a}. The first equilibrium (Fig. \ref{fig1a}) has nested
magnetic surfaces and the others
 (Figs. \ref{fig2a}, \ref{fig3a})  non-nested ones; specifically the second equilibrium
  consists of a couple of magnetic
 lobes with magnetic axes orientated  perpendicular to the axis of symmetry
 (Fig. \ref{fig2a}) and the  third  one has three lobes with magnetic axes
 orientated parallel to the axis of symmetry (Fig. \ref{fig3a}). In the
 second and third equilibria the plasma flows parallel to the magnetic
 field ($\gamma_1 = 0$).  The single magnetic axis of the first equilibrium
 is located at $\rho\approx 1.002$ and its position remains nearly unaffected
 when the values of the flow parameters $M_0^2$ and $\gamma_1$ change.
\begin{table}[t]
\centering
\begin{tabular}{|c|cccc|}
\hline Parameters &$\alpha_1$ &$\beta_1$& $\gamma_1$&$\alpha$ \\
[0.5ex] \hline\hline Fig. 2 & 5.3 & 1.3 & 1.3& 5.0  \\ \hline Fig.
3 &1.5 &1.5 &0 &3.5  \\ \hline Fig. 4 & 7.5 &7.0& 0& 6.75 \\ [1ex]
\hline
\end{tabular}
\caption{Values of the parameters in ansatz (\ref{pc1}-\ref{pc3})
for the equilibria of Figs. \ref{fig1a}-\ref{fig3a}.} \label{tab1}
\end{table}


\begin{figure}[ht!]
\centerline{\mbox {\epsfxsize=10.cm \epsfysize=8.cm
\epsfbox{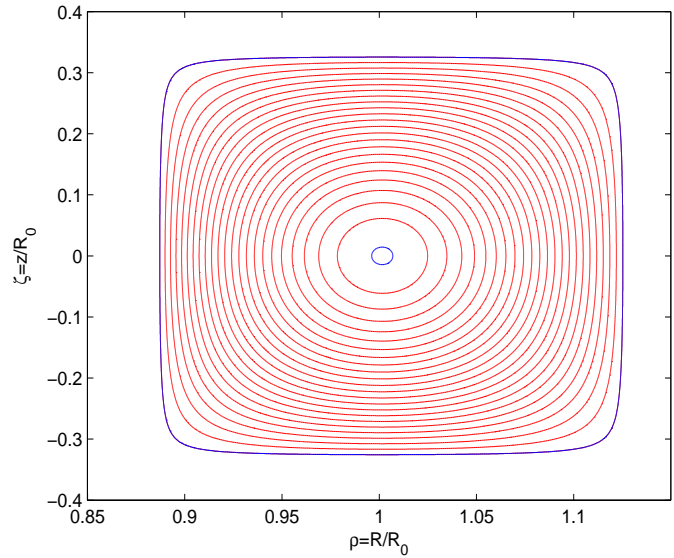}}} \caption[]{The first equilibrium with nested
magnetic surfaces described in the text. The boundary, shown in
blue, corresponds to the flux value of $u_{b}=0$, while on the
magnetic axis we have $u_{a}=-0.9375$.} \label{fig1a}
\end{figure}

\begin{figure}[ht!]
\centerline{\mbox {\epsfxsize=10.cm \epsfysize=8.cm
\epsfbox{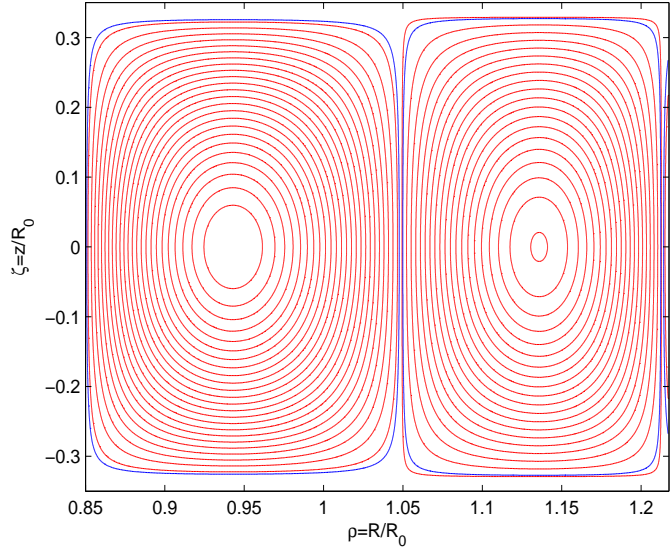}}} \caption[]{The second equilibrium with
non-nested magnetic surfaces described in the text. The boundary of the
magnetic lobes, shown in blue, corresponds to the flux value of $u_{b}=0$,
while on the left magnetic axis we have $u_{aleft}=-0.94$ and on the
right one $u_{aright}=0.9$.} \label{fig2a}
\end{figure}

\begin{figure}[ht!]
\centerline{\mbox {\epsfxsize=10.cm \epsfysize=8.cm
\epsfbox{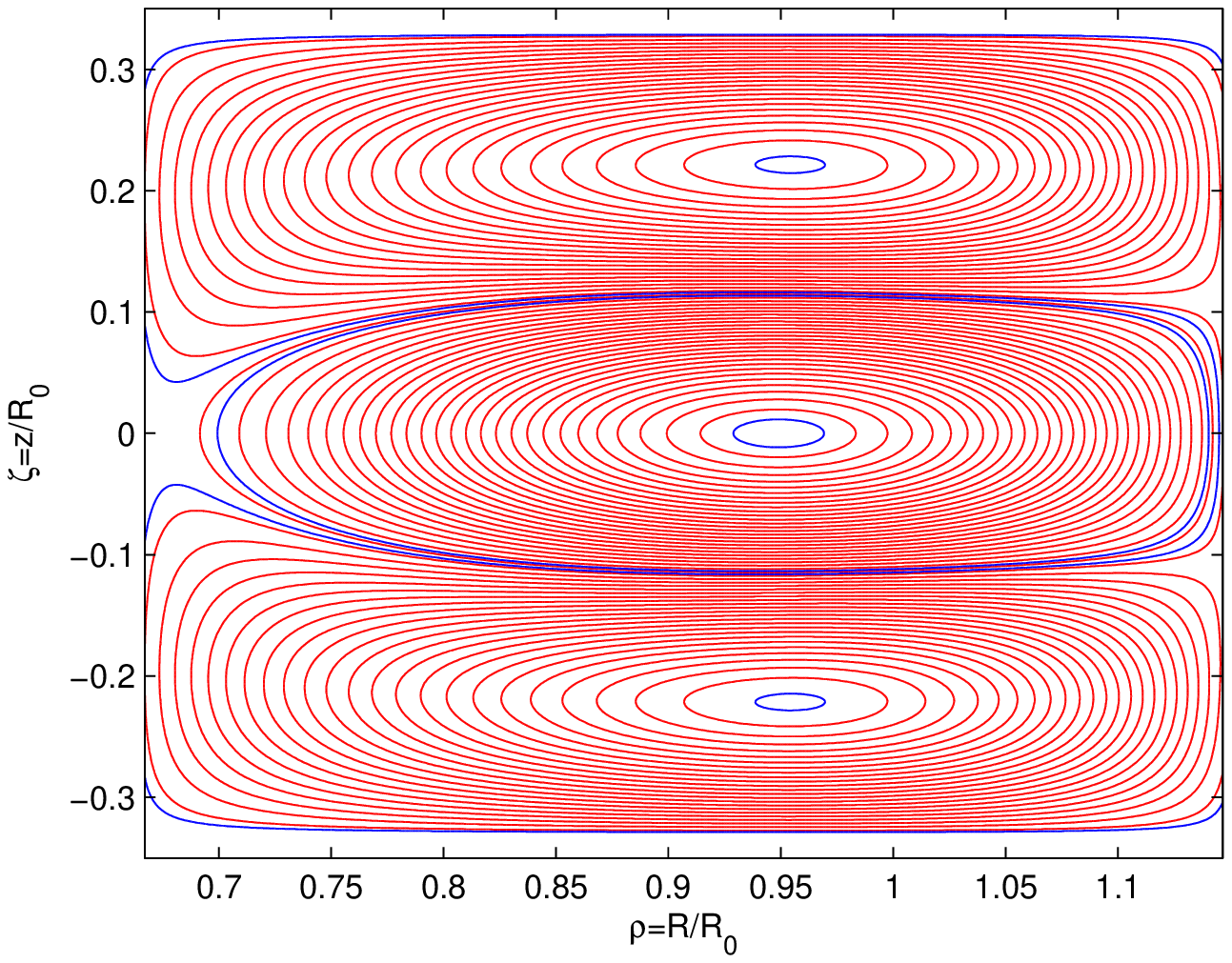}}} \caption[]{The third equilibrium with
non-nested magnetic surfaces described in the text. The boundary of the
magnetic lobes, shown in blue, corresponds to the flux value of $u_{b}=0$, while
at the magnetic axes, as measured from top to bottom we have
$u_{a1}=u_{a3}=0.995$ and $u_{a2}=-1.12$.}
\label{fig3a}
\end{figure}

We have examined the impact of the flow on the toroidal current density of equilibria
with nested magnetic surfaces and found that the non parallel component of the flow
 can result in current reversal in the region of the
magnetic axis.  This is  illustrated in Fig. \ref{fig11a} corresponding to the peaked on-axis choice (\ref{mach1}) with $n=2$.  The continuous curve
 therein
represents a static equilibrium
($M_0^2=\gamma_1=0$) for which current reversal is not possible as already mentioned
in section 1. Central current reversal occurs at a critical value of $(\gamma_1)_c=3.0$,
nearly irrespective of the value of the parallel-flow
parameter $M_0^2$ [see Eqs. (\ref{decomp}) and (\ref{mach1})]. This result is consistent with that of \cite{po} where
 purely toroidal flow
was considered. To check that the parallel flow has no appreciable impact on the central current reversal we also employed the peaked off-axis flow profile (\ref{mach}) with $m=n=2$. The results respective to those of Fig. \ref{fig11a} are shown in Fig. \ref{fig11b}. Apparently the current density profiles therein are identical with those of Fig. \ref{fig11a} in the close vicinity of the magnetic axis. The off-axis flow, though, makes the region of current reversal narrower.

The pressure of the equilibrium of Fig. \ref{fig1a} monotonically decreases from the magnetic axis to
the plasma boundary as can be seen in Fig. \ref{fig5a}.
 In addition, the safety factor monotonically increasing
from the magnetic axis to the plasma boundary is shown in Fig. \ref{fig7a}.

Finally, it is noted that current reversal can occur in equilibria with non-nested magnetic surfaces even in the absence of flow \cite{mart}-\cite{maro}.
 \begin{figure}[ht!]
\centerline{\mbox {\epsfxsize=10.cm \epsfysize=8.cm
\epsfbox{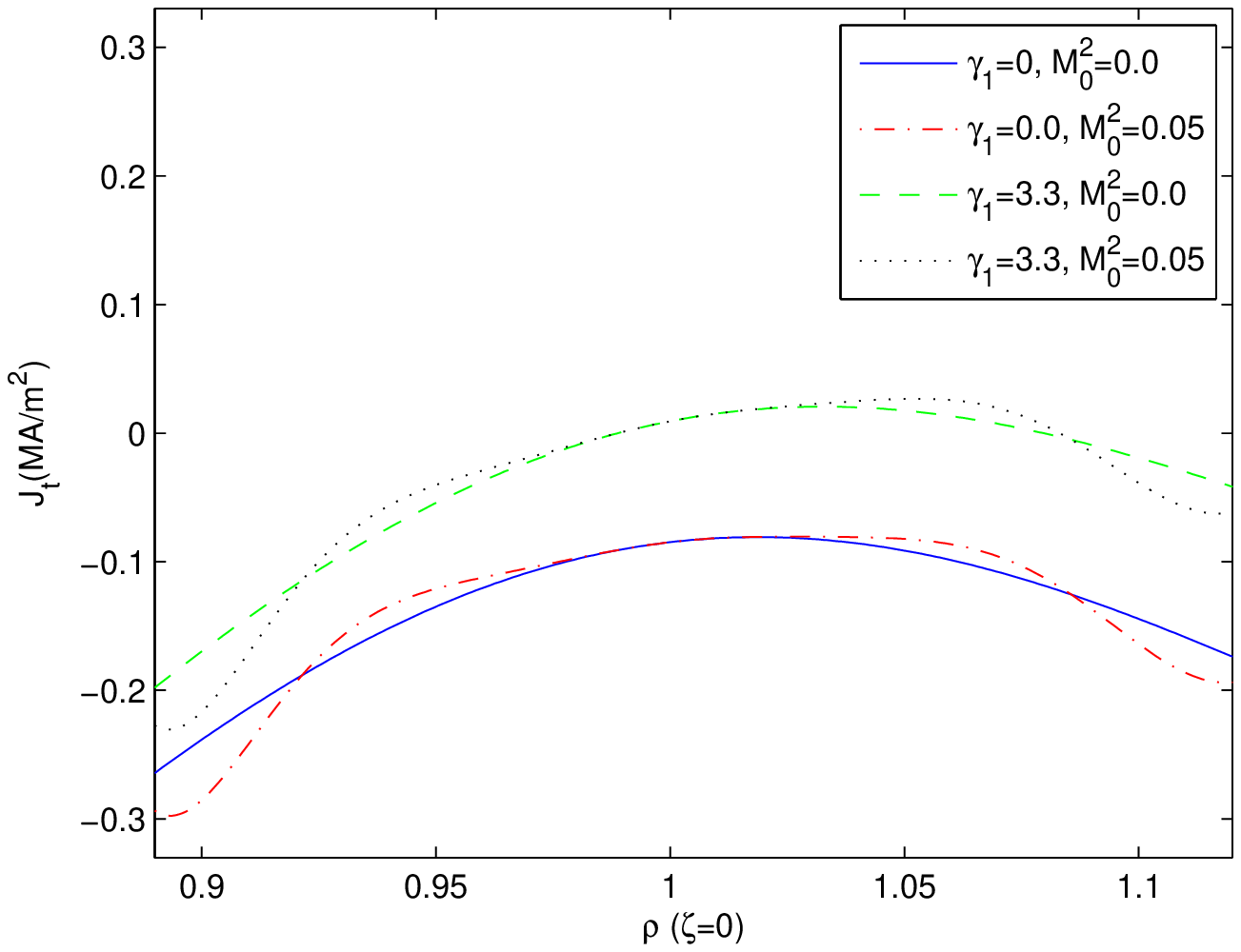}}} \caption[]{Profiles of the the toroidal current
density on the middle-plane $\zeta=0$ for the equilibrium with nested magnetic surfaces  (Fig. \ref{fig1a}),
for various values of the  flow parameters $M_0^2$ and $\gamma_{1}$. This refers to the peaked-on-axis choice of Eq. (\ref{mach1}).
For the green and black curves, we found that on the magnetic axis
the value $J_{ta}\simeq 9700 A/m^2 >0$ and thus indeed current density reversal
occurs thereon due to the non-parallel flow (parameter $\gamma_1$). The parallel component of the flow associated with $M_0^2$ hardly affects this reversal.
}
\label{fig11a}
\end{figure}

 \begin{figure}[ht!]
\centerline{\mbox {\epsfxsize=10.cm \epsfysize=8.cm
\epsfbox{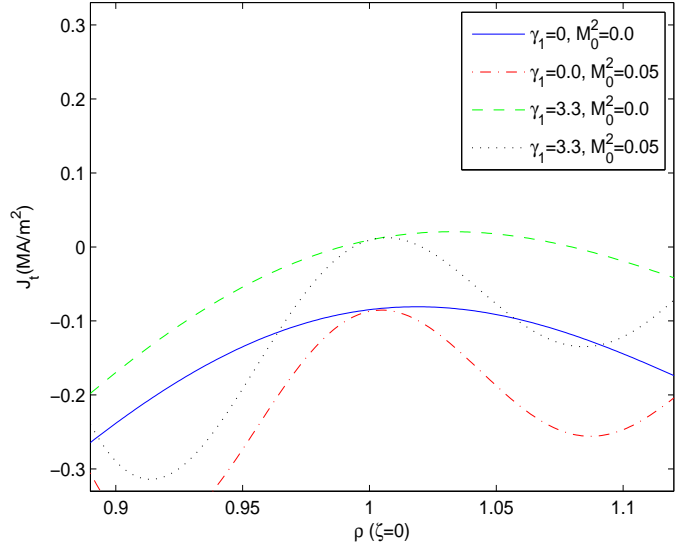}}} \caption[]{Profiles of the the toroidal current
density on the middle-plane $\zeta=0$ for the equilibrium with nested magnetic surfaces  (Fig. \ref{fig1a}),
for various values of the  flow parameters $M_0^2$ and $\gamma_{1}$.
 This refers to the peaked-off-axis choice of Eq. (\ref{mach}). }
\label{fig11b}
\end{figure}

\begin{figure}[ht!]
\centerline{\mbox {\epsfxsize=10.cm \epsfysize=8.cm
\epsfbox{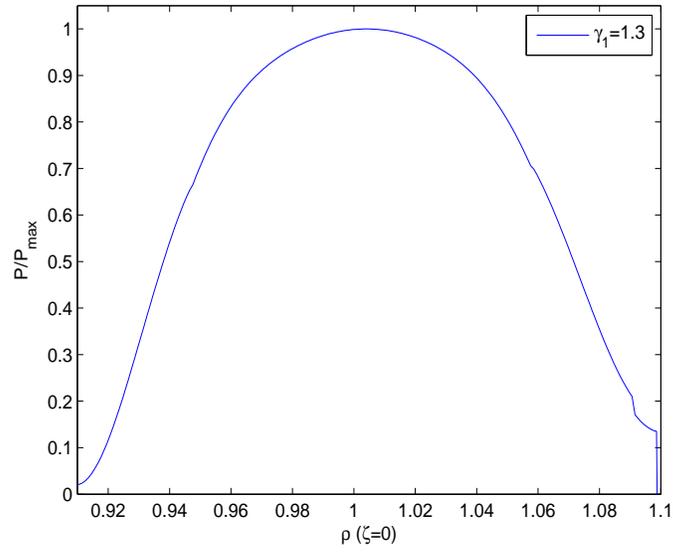}}} \caption[]{ The pressure for the
equilibrium of Fig. \ref{fig1a}}
\label{fig5a}
\end{figure}
\begin{figure}[ht!]
\centerline{\mbox {\epsfxsize=10.cm \epsfysize=8.cm
\epsfbox{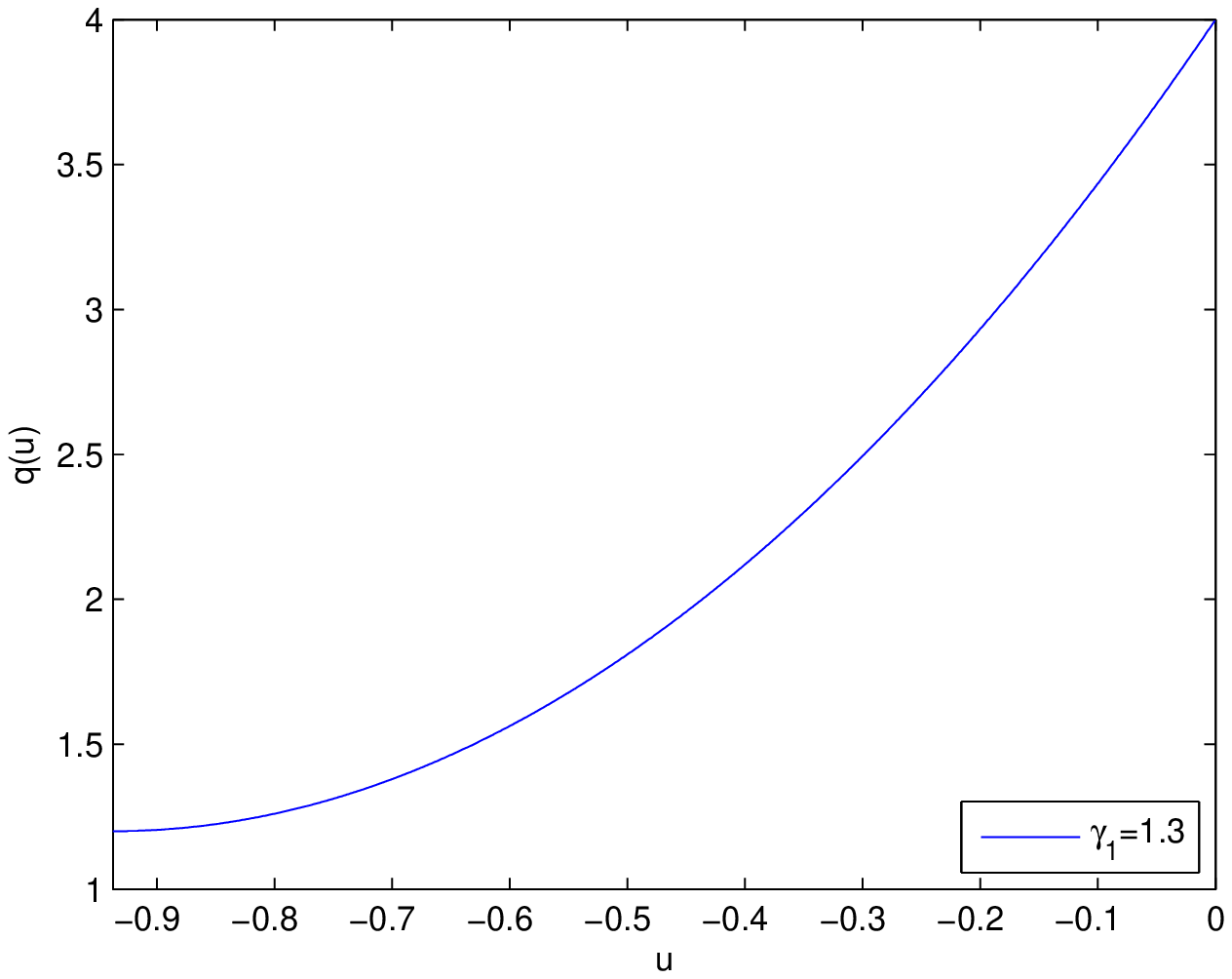}}} \caption[]{ The safety factor
for the equilibrium of Fig. \ref{fig1a}.}
\label{fig7a}
\end{figure}

%
\section{Summary}
Solving the GGS equation (\ref{gsh1}) analytically we have
constructed equilibria with incompressible flow, reversed toroidal
current density and either nested or non-nested magnetic surfaces. The
non-parallel flow results in normal  equilibria
with nested magnetic surfaces with central  current-reversal and
monotonically varying pressure profiles.  This reversal gets stronger as  the non-parallel flow parameter
$\gamma_1$ in Eq. (\ref{pc3}) takes larger values and is nearly unaffected by the parallel component of the flow.
 Finally, it is noted that the stability of equilibria with non parallel
 flow remains a tough unsolved problem reflecting the absence of stability
  criteria even in the framework of hydrodynamics.
%
%
%
\section*{Aknowledgments}
This work has been carried out within the framework of the
EUROfusion Consortium and has received funding from  the
National Programme for the Controlled Thermonuclear Fusion,
Hellenic Republic.
 The views and
opinions expressed herein do not necessarily reflect those of the
European Commission.
%
\section*{Appendix: Solution of the homogeneous Eq. (\ref{bessl})}
Since  Eq. (\ref{bessl}), put in canonical form,  has a   regular
singular point at $\rho=0$ it can be solved  by the Frobenius
method of infinite convergent series around this point, e.g. \cite{fong}. Here we
will employ this method slightly modified. Accordingly
we try a solution of Eq. (\ref{bessl}) of the form \bea \label{a1}
U_{1}:=\sum_{n=0}^{\infty}u_{n}\rho^{n} \eea We obtain $u_{0}=0$,
$u_{1}$ is a freely specified constant which is hereafter,
according to standard analysis \cite{fong}, is set equal to
$u_{1}=0.001$, $u_{2}=0$, $u_{3}=-\mu^{2}u_{1}/8$, $u_{4}=0$ and
\bea \label{rcr1}
u_{n}=-\frac{1}{(n^{2}-1)}[\mu^{2}u_{n-2}+\lambda^{4}u_{n-4}],\; \;
(n\geq 5) \eea
where $\lambda^{2}=\mu^{2}+\nu^{2}$.
We hereafter call this solution $J_{1}(\rho;\mu,\nu)$.
For sufficiently small values of the parameter $\mu$,
$J_{1}$ has the usual oscillatory behaviour of the
standard Bessel function.
More specifically we have found, using extensive numerical tests,
that for every $0<\rho_{1}<\rho_{2}$ there exist numbers
$\epsilon_{1},\; \epsilon_{2}$ with $0<\epsilon_{2}<\epsilon_{1}$
and an integer $n_{0}$ such that $|U_{1}(\rho_{1})|<\epsilon_{1}$
and  $|U_{1}(\rho_{2})|<\epsilon_{2}$ provided that the number of
terms $n_{u}$ used in the series expansion of $U_{1}$ satisfies
$n_{u}\geq n_{0}$. For computational purposes we have retained
terms in the series expansion of up to $N_{s}=100$.

The second independent solution of Eq. (\ref{bessl})will be of the form
 \bea
\label{a2} U_{2}=U_{1}(\rho)ln(\rho)+\frac{b_{-1}}{\rho}+
\sum_{n=0}^{\infty}b_{n}\rho^{n} \eea and obtain $b_{0}=0$,
$b_{-1}=-1/\mu^{2}$. Also,  $b_{1}$ is a free coefficient which
is set, according to the analysis in \cite{fong} equal to
$b_{1}=-0.001$, $b_{2}=0$, \\
$8b_{3}=\left(\frac{3\mu^{2}}{8}+\frac{\lambda^{4}}{\mu^{2}}-b_{1}\mu^{2}\right)$,
$b_{4}=0$ and \bea \label{rcr2}
(n^{2}-1)b_{n}=-2nu_{n}-\mu^{2}b_{n-2}-\lambda^{4}b_{n-4},\; \; (n\geq
5) \eea
We call this solution hereafter $Y_{1}(\rho;\mu,\nu)$.

Since the closest singular point of (\ref{bessl}) to  $\rho=0$ is
infinity,  solutions $J_1(\rho;\mu,\nu)$ and $Y_1(\rho;\mu,\nu)$
converge for any finite value of $\rho$.

\end{document}